\documentclass[aps,11pt,amsmath,amssymb,nofootinbib,a4paper,showpacs]{revtex4}%twocolumn
\usepackage[breaklinks]{hyperref}
\usepackage[latin1]{inputenc}
\usepackage{graphicx}
\usepackage{color}
\usepackage{amsfonts,amsthm}
\usepackage{bm,bbm}
\usepackage[OT2,OT1]{fontenc}
\newcommand\cyr{%
\renewcommand\rmdefault{wncyr}%
\renewcommand\sfdefault{wncyss}%
\renewcommand\encodingdefault{OT2}%
\normalfont
\selectfont}
\DeclareTextFontCommand{\textcyr}{\cyr}

\begin{document}
\title{Spacetime as a quantum many-body system}
\author{{\bf Daniele Oriti}}
\affiliation{Max Planck Institute for Gravitational Physics (Albert Einstein Institute) \\ Am Muehlenberg 1, D-14476 Potsdam-Golm, Germany, EU \\ daniele.oriti@aei.mpg.de}
%\date{\small July 30, 2014}
\begin{abstract}
Quantum gravity has become a fertile interface between gravitational physics and quantum many-body physics, with its double goal of identifying the microscopic constituents of the universe and their fundamental dynamics, and of understanding their collective properties and how spacetime and geometry themselves emerge from them at macroscopic scales. In this brief contribution, we outline the problem of quantum gravity from this emergent spacetime perspective, and discuss some examples in which ideas and methods from quantum many-body systems have found a central role in quantum gravity research. 
\end{abstract}
%\pacs{04.60.-m, 98.80.Qc}
\maketitle
\section{The problem of quantum gravity and its evolution}
The construction of a complete theory of quantum gravity remains an open issue, despite decades of activities and important progress \cite{QGhistory}. These have led to a number of approaches \cite{QGbook} and many fascinating suggestions for a radically new basis for understanding the world, which remain, however, still incomplete and not directly tested by observations. To communicate this variety and fascination is the first goal of this contribution. Over the last two decades, the very perspective on the problem has changed, as a result of new insights obtained in different quantum gravity approaches and some surprises found along the way. The new perspective brings the conceptual framework of quantum gravity closer to that of condensed matter theory, with a consequent import of mathematical tools and continuous exchanges between these two areas. To explain this change in perspective is the second goal of this contribution. This evolution is apparent in several corners of current research in quantum gravity, and very concrete: a number of quantum many-body systems have been studied in order to obtain insights for quantum gravity research, quite a few others have been shown to be in very direct correspondence with quantum gravitational phenomena, and quantum gravity formalisms themselves have developed in such a way that the have a closer resemblance to quantum many-body systems. This contribution will therefore present also a quick survey of this multi-facets relation between quantum gravity and quantum many-body systems. The vast amount of literature that has accumulated in recent years on this interface, let alone quantum gravity per se, exceeds greatly our own competence, and certainly the space of this contribution. This will therefore be limited to a summary of interesting developments, without entering into any detail, and it is only meant to be an invitation to dwelling more into this subject and a pointer to the relevant literature.

\subsection{Quantum gravity as quantised GR}
The main attitude towards the problem of quantum gravity has been for many years \cite{QGhistory} (and still is, in many corners) the straightforward one applied to all other fundamental interactions: start from the best classical description we have of the interaction and quantize it, i.e. apply to the corresponding classical theory well-tested techniques for turning it into a quantum theory. The starting point is of course General Relativity (GR), the modern description of gravitational phenomena. The variety of quantization methods (canonical quantization schemes, geometric quantization, path integral methods, etc) already led to a variety of quantum gravity formalisms. General Relativity is however a very complicated theory at the mathematical level (e.g. it is highly non-linear and it is invariant under spacetime diffeomorphisms, a notoriously involved symmetry group) and this already suggests that progress should be expected to be difficult. Severe challenges come also from the conceptual revolution that GR forces into our understanding of the world. The gravitational field is identified with spacetime geometry, on which all other interactions rely, and this in turn determines the causal structure of all physical events. Spacetime itself becomes therefore a dynamical, physical system, and we have no experience in dealing with physical processes in presence of a dynamical spacetime, at the quantum level. Also diffeomorphism invariance expresses this need to define localization of physical objects in time and space in relational terms, i.e. in terms of one another, without relying on any fixed notion of time and space. This immediately calls into question the notions of time evolution, conservation of probabilities and unitarity of usual quantum field theory, together with the causality restrictions that enter its very foundation. 

Some quantum gravity approaches take onboard these lessons from GR and try to apply quantization techniques to the full gravitational field, i.e. to spacetime geometry itself: canonical quantum gravity and its modern incarnation Loop Quantum Gravity \cite{LQG}; path integral formulations of quantum GR and their modern evolution mostly based on lattice structures (e.g. causal dynamical triangulations \cite{CDT} and spin foam models \cite{SF}, or group field theories \cite{GFT}) or other forms of discrete substrata like Causal Set Theory \cite{CS}; other strategies to quantize spacetime structures, like non-commutative geometry \cite{NCG}. 

Researchers coming from a particle physics tradition have instead often relied on different way of tackling the problem of quantizing the gravitational field. The basic intuition comes from splitting the spacetime geometry into a (usually flat) fixed background and a dynamical part, which is then quantized by standard quantum field theory methods. The result is a dynamical quantum theory of \lq gravitons\rq, quanta of the gravitational field defined with respect to the fixed background spacetime. Prima facie, this procedure runs contrary to the main lessons of GR, but it allows to take on board most techniques of the usual quantum field theory formulation of other fundamental interactions. However, one could expect that the quantum field theory of gravitons, if taken beyond the perturbative regime to account for all the non-linearities and non-perturbative features of GR, and its full symmetry content, will in the end lead as well to a theory of the whole quantum spacetime giving answers to the same basic issues that GR-based approaches face. If this is achieved, the fixed background geometry that enters the initial formulation would be recognised as an auxiliary tool with no physical content, as in standard background-field methods in QFT. The fact that going beyond the perturbative regime is an absolute necessity is further confirmed by the failure of the perturbative quantum field theory of gravitons to be consistent, having been shown to be non-renormalizable. 

The non-perturbative route to a quantum field theory of the gravitational field is for example taken by the asymptotic safety approach \cite{AS}. String theory \cite{ST} was born as an enrichment of the perturbative approach to quantum gravity in terms of gravitons, by adding to them more degrees of freedom (an infinity of them), naturally resulting from the step from point-like objects to extended, string-lke entities propagating on the same fixed background; it has also revealed a vast number of non-perturbative phenomena which involve extended spacetime configurations and quantum aspects of the gravitational field, and a new understanding of quantum field theories themselves (but not yet a clear, if tentative, picture of the fundamental nature of spacetime itself). 

All the modern approaches to quantum gravity show that the straightforward perspective on quantum gravity as \lq quantised GR\rq, in its many versions, can lead to important results and insights. 

At the same time, a slow change in perspective has taken place over the last (15 or so) years, with the result that the very problem of quantum gravity is now understood from a different angle, and many quantum gravity researchers are now inclined to take a more radical standpoint. 

This has also led to a closer relation between this research area and condensed matter theory (as well as quantum information). 

\subsection{The emergent spacetime scenario}
The first hint that something more radical is in store came already in the seventies (of the last century) with the development of black hole thermodynamics \cite{BHthermo}. The discovery, within the semi-classical, very conservative framework of quantum field theory in curved backgrounds, that a finite entropy can be associated to a black hole and that the same can emit thermal radiation, leading to a consistent (if rather mysterious) set of thermodynamical laws, had tremendous implications. Indeed, these laws simply cannot be made sense of within classical physics. The finite entropy, in particular, strongly hints at some underlying discrete microstructure of spacetime, since black holes are in the end particular configurations of it. The task of quantum gravity approaches became (also) to unravel such discrete microstructure and compute this entropy from first principles. In the ensuing decades, semi-classical black hole physics have turned into an unrivalled source of radical suggestions \cite{BHthermo2}. the end result of black hole evaporation remains a mystery (a Planck-sized exotic object? nothing?) just as a mystery remains the fate of the information stored in the evaporating black holes (forever lost? somehow encoded in the outgoing radiation? stored in the exotic object that is left over?), but any proposed solution to these mysteries calls for relinquishing at least one of the sacred ingredients of modern theoretical physics, be it locality, the equivalence principle, unitarity, causality, quantum monogamy or no-cloning, or the real dimensionality of spacetime. And this is {\it in addition} to the continuum spacetime structure. 

Related to this, a number of developments, starting from Jacobson's derivation of the Einstein's equations as an equation of state, have strongly supported the idea that GR itself, and Einstein's equations of motion, can be seen as the macroscopic thermodynamic (or hydrodynamic, depending on the specific framework) description of an underlying set of non-geometric (thus not directly \lq gravitational\rq) degrees of freedom \cite{GRthermo}.

More recently, the main drivers of such change in perspective have been the same approaches to quantum gravity mentioned above. Indeed, the more they were unraveling about the fundamental nature of gravity and spacetime, the more the new insights pushed for a different point of view and for the use of new tools.What happened is that all these different approaches have either identified explicitly or at least strongly suggested a set of fundamental degrees of freedom underlying spacetime which are not simply quantised continuum fields and are nor directly geometric, nor characterised in continuum spatiotemporal manner. Let us mention a few examples. 

Loop quantum gravity has proceeded far enough as to identify a candidate Hilbert space of quantum states for the gravitational field; such states admit a characterization in terms of \lq generalised continuum fields\rq , under appropriate conditions, but still they are labeled by purely combinatorial and algebraic structures: graphs labeled by data from the representation theory of $SU(2)$ (or the Lorentz group), defining so-called \lq spin networks\rq. Moreover, the continuum interpretation is possible only in a limit; each state can at best, i.e. under additional semi-classical and regularity conditions, be put in correspondence with a piece-wise flat geometry \cite{SN-geometry}, as in lattice gravity, rather singular from the continuum GR perspective. 

Group field theories  \cite{GFT2nd} and spin foam models \cite{SF} have the same type of quantum states. 

In fact, the same type of singular configurations are used as basic entities (even if mostly interpreted just as regularization tools) in lattice approaches to quantum gravity, like (causal) dynamical triangulations \cite{CDT}. Thus, a generic configuration of fundamental degrees of freedom in these formalisms will not correspond to anything resembling a nice continuum spacetime or geometry, even if a classical limit is taken. 

Causal set theory is another formalism in which purely discrete structures (partially ordered sets), even though strongly motivated from continuum geometric considerations, replace entirely the continuum and geometric structures on which GR is based, and in which their generic configurations do not admit any immediate interpretation in terms of spatiotemporal notions like distances or even dimensionality, which have to be instead reconstructed with laborious procedures \cite{CS}.  

Of course, the actual distance between the quantum configurations at the basis of a given approach and the usual notions of spacetime and geometry can vary, depending on the specific version of the formalism adopted and additional conditions that may be imposed on them. 

From the other side of the spectrum of quantum gravity approaches, also string theory has produced a large number of challenges to the usual notions of spacetime and geometry, and hints that the fundamental degrees of freedom of the theory (whatever they are) will not be anywhere close to straightforward quantised continuum fields \cite{STemergence}. T-dualities indicate that string theories on a spacetime in which one spatial direction is compactified to a small radius can actually be equivalent to string theories in which the same compact dimension has a very large radius; mirror symmetry shows that some formulations of string theory on a given topology can be physically equivalent to formulations on an entirely different topology; the very notion of spacetime dimension is dynamical, since it can be \lq exchanged\rq with a different field content and internal symmetry, while the background \lq geometry\rq on which strings propagate and interact can be very exotic, going beyond the usual notion of Riemannian geometry of GR, but also beyond the already exotic non-commutative but associative geometries used in other approaches to quantum gravity \cite{string-geometry}. 

Therefore we have by now many reasons to believe that spacetime, at the fundamental level, \lq dissolves\rq ~into a new set of quantum entities, call them \lq atoms of space\rq , of no direct gravitational or spatiotemporal or geometric interpretation, and from which it has to re-emerge, together with the usual notion of geometry and gravity, in some limit and under suitable approximations \cite{emergence}. 

The detailed nature of the fundamental entities, to the extent in which it is identified, varies greatly across different quantum gravity formalisms, just as their \lq weirdness\rq as compared to the usual notions of continuum geometry, spacetime, fields, and so on. At the same time, different formalisms enjoy a number of mutual relationships, both conceptual and mathematical, and their different suggestions need not be seen as incompatible with one another, nor, necessarily, with the traditional view of quantum gravity as \lq quantised GR\rq. However, the fact remains that some more conceptual flexibility is called for, and a new set of tools is required, to tackle the new issue of the emergence of spacetime from this new type of entities. Let us dwell more into the novel aspects of this situation.  

The change of fundamental degrees of freedom implies that a classical approximation is not enough to recover GR, which is not only a classical theory but also based on continuum spacetime and geometry, which have to emerge first. Given that such emergent quantities have a continuum character and sustain a field theory framework with its infinite number of degrees of freedom it is reasonable to expect that such continuum approximation (independent, we repeat, from the classical approximation) requires the control over a very large number of the fundamental entities. 

The key tool to achieve such control and to define the continuum limit is the renormalization group, which has to help us to map the macroscopic phase diagram of any given quantum gravity model. In particular, we should seek to identify the phase where the continuum {\it geometric} characterization of the emergent physics is possible, and the phase transition separating it from non-geometric phases, which cannot describe our world. In any geometric phase so identified, we also have to be able to characterize in detail the collective dynamics of the atoms of space and any emergent property coming out of it, translating it in the language of GR and effective field theories, as needed to extract phenomenological predictions. 

This list of tasks is common to all approaches to quantum gravity.  
It is also a more abstract and, admittedly, exotic analogue of the list of tasks that any theory of \lq real\rq atomic, many-body systems has to undertake to go from the identification of the fundamental constituents to the characterization of their collective, macroscopic properties. In this sense, quantum gravity is now forced to take onboard the many lessons and tools of condensed matter theory, solid state physics and, more generally, many-body quantum theory in order to make further progress.

This change in perspective comes with a new set of questions and fascinating possibilities, actively explored. Among them: is geometry emergent from entanglement (of the fundamental \lq atoms of space\rq)? what is the geometric, gravitational counterpart of other quantum properties of the underlying atoms of space? should we drop basic features of effective field theories, like locality or unitarity? what about the very notion of separation of scales, which is based on the existence of a given geometry and it is what suggests that quantum gravity effects should only become relevant at microscopic, Planck scales? if we drop this, and more generally if the very notion of spacetime is emergent, what prevents phenomena that we usually consider macroscopic, large-scale aspects of the world (e.g. the cosmological constant, dark energy and the current accelerated expansion of the universe) to be really of quantum gravity origin? The list could go on for quite a while. Rather than listing possibilities, which we would have no way to discuss properly in this contribution, we move on instead to give a few examples of many-body systems which offer insights for quantum gravity research, that are directly connected to (quantum) gravitational phenomena, or that {\it are} themselves candidate quantum gravity systems.   

\section{The interface between quantum gravity and many-body systems: some examples}
\subsection{Analogue gravity in condensed matter systems} 
The first examples of many-body systems of interest for quantum gravity are actually standard condensed matter systems, of great interest as such, and are instances of so-called \lq analogue gravity models\rq . A very comprehensive account can be found in \cite{analogue}. 

The basic fact about all of them is surprising as it is interesting: propagating excitations over given macroscopic (more or less special) background configurations of the system are described by (quantum) field theories on an emergent curved geometry, whose metric is a function of macroscopic collective variables. The interesting fact is both that these excitations couple to an effective curved metric, rather than the flat one associated to the laboratory and the underlying atomic system, and that this metric is general enough to include configurations mimicking black hole horizons or cosmological spacetimes. Beside the theoretical interest as models for quantum effects on curved spacetimes (but with modified dispersion relations and Lorentz breaking at high energies) and the insights these systems can offer on quantum gravity and the nature of spacetime, they have also raised the hope of actual observation of such effects in a controlled laboratory environment. 

In fact, the simplest example of analogue gravity is given by ordinary fluids: the velocity potential for an acoustic perturbation of any barotropic and inviscid fluid with irrotational flow is governed by an equation that can be recast in the form of a D'Alambertian equation for a minimally coupled, massless (real) scalar field propagating on a curved geometry with a metric that is an algebraic function of the fluid density and the flow velocity, and the local speed of sound. The emergent geometry is general enough to reproduce features of black hole horizons and their surface gravity, as well as many of their causal properties. Also, depending on the details of the fluid flow, it may reproduce various geometries of cosmological interest, and it may be used, for example, to simulate cosmological particle production in an expanding universe. 

This is the simplest example of a long list of condensed matter and solid state systems where this type of behaviour occurs. The most studied ones are Bose condensates and fermionic superfluids, which show an even richer emergent gravitational-like physics, thanks to their quantum coherence and low speed of sound, as well as gravity analogues in quantum optics. More recently, analogue gravity models have been developed using graphene, also showing a rich set of emergent phenomena. 

The range of topics, at the interface between gravitational physics and quantum field theory, that analogue gravity systems have allowed to investigate from a new angle is impossible to even summarize here (to give just one more example, they have suggested new ways of tackling the cosmological constant problem \cite{analogue-cosmologicalconstant}), so we refer again to the literature \cite{analogue}. These models have also been quite influential in quantum gravity research even beyond the actual results obtained, since they have provides examples of the emergence of gravitational phenomena and effective spacetimes with non-trivial geometry from systems where gravity played no role and geometry was trivial, and thus realised to some extent the idea of geometry from non-geometry. Therefore, they have provided further support to the whole idea that spacetime and geometry could be emergent notions even at the fundamental level of quantum gravity, beyond simple analogy \cite{hu, GFTfluid}. 

Having said this, it is also to be noted that these systems also have strong limitations from the point of view of fundamental quantum gravity: the emergent matter fields propagating on the effective curved geometry do not have the right type of backreaction on it, i.e. they do not \lq gravitate\rq as prescribed by GR; moreover, and perhaps most importantly, it has proven very hard to reproduce in such many-body (quantum) systems the gravitational {\it dynamics} for the effective metric, i.e. some possibly modified version of the GR equations, characterized by a general covariance that it seems very hard to mimic, even approximately. Many-body quantum physics systems have started playing a prominent role, however, also in actual quantum gravity formalisms and true gravitational theories, to which we now turn.  

\subsection{AdS/CFT correspondence and geometry from entanglement}
A most extensive application of many-body systems in quantum gravity has stemmed from the establishment of the AdS/CFT correspondence \cite{AdS/CFT} as one of the most active and fertile research directions in the theory of fundamental interactions, starting from string theory but with an influence extending well beyond it. 

In brief, the AdS/CFT correspondence is a map between a conformal field theory (CFT) living on a flat d-dimensional Minkowski spacetime and a gravitational theory living on a (d+1)-dimensional curved spacetime with a metric approximating the Anti-DeSitter (AdS) one close to its flat boundary, where it reduces to the Minkowski metric on which the CFT depends. The map implies that the partition function of the (quantum) gravitational theory, with the mentioned boundary conditions, its observables and Hilbert space of states have a precise counterpart in the partition function and observables of the boundary CFT: whatever one computes within one theory, one is implicitly performing some correspondent computation within the other. The correspondence has been first proposed to hold between one specific CFT, which is also a specific gauge theory, i.e. SU(N) Yang-Mills theory with N=4 supersymmetry in 4 spacetime dimensions, and a rather exotic gravitational theory, i.e. IIB superstring theory living on a 10-dimensional spacetime in which 5 dimensions are compactified to a 5-sphere (and play basically the role of \lq internal\rq dimensions) and the others have AdS boundary conditions. However, it has since been extended to quite a few other CFTs (with varying degree of supersymmetry and with different gauge symmetries) and to other gravitational theories (including other formulations of string theory, but also many other results concerning generic semi-classical gravity and quantum field theories in curved spacetimes), and to different spacetiime dimensions, including lower dimensional contexts. 

One interesting example is the map of the observable properties of a gas of bosons, fermions and scalars at thermal equilibrium in the CFT to the (semi-classical) observable properties of a black hole living inside an AdS spacetime, with the equilibrium temperature of the gas being the Hawking radiation temperature of the black hole. Another recent and very simple example (possibly the simplest example of AdS/CFT correspondence) is the SYK model (which has been generalised in various ways), a statistical system of Majorana fermions in $0+1$ dimensions with random interactions, which is the only known system that is simultaneously solvable at strong coupling, maximally chaotic and with an emergent conformal symmetry \cite{SYK}; the AdS/CFT correspondence suggests that this model relates to a black hole in a 2-dimensional AdS spacetime; moreover, its relation to quantum gravity is strengthened by the fact that its quantum interaction processes can be mapped to the interaction processes of random tensor models \cite{tensors, tensors-SYK}, themselves models of quantum gravity closely related to group field theories.  

The conjecture (still unproven) is that the map defines an exact AdS/CFT {\it duality}, i.e. the two theories are strictly equivalent despite being so different and living in different spacetimes (with different dimensions!). Some researchers, on the other hand, think of it more as an {\it approximate} map, involving coarse-graining of degrees of freedom or holding only in appropriate limits of the two theories (most of the evidence supporting it, we remark, concerns only semi-classical gravity), or an exact one but holding only for a subset of observables and configurations. 

The hope, supported by some circumstantial evidence, is that this correspondence is only the most evident example of a much more general one between gravity and gauge theory tout court (thus holding for any gauge theory, with any choice of boundary conditions, in any dimension, etc). 

The details of what is known and what is not in this context do not concern us too much. The general idea does. The AdS/CFT correspondence gives us examples of theories in which properties of a standard many-body quantum system, the CFT in a flat spacetime, which a priori does not know about gravity (classical or quantum) or curved geometries, somehow produce full-blown gravitational effects and can be translated in terms of curved {\it dynamical} geometries. Of course, the interest is even greater if one believes in an exact duality that extends beyond the AdS boundary conditions, since it would imply that quantum gravity is actually {\it defined} by some specific quantum many-body system. But even leaving aside this belief, it remains the fact that non-trivial geometric aspects of the world can emerge from quantum mechanical properties of a many-body system living in a flat spacetime. Inspired by and taking advantage of the concrete testbed of the AdS/CFT correspondence, quantum gravity researchers have then explored a variety of possibilities. 

One of the most fascinating is that geometry can actually be entirely characterized in terms of the quantum information-theoretic properties of a many-body system, in particular entanglement \cite{entanglement-geometry}. To start with, this has prompted the application of powerful tools from quantum information theory and quantum condensed matter systems to gravitational physics, the most notable example being tensor network techniques \cite{tensor-networks}. The idea of \lq geometry from entanglement\rq has already found support in a number of results. For example, the entanglement entropy between a region $A$ on the flat boundary of an AdS space and the rest of such boundary, computed within a simple CFT,  is proportional to the area of the minimal surface inside the bulk AdS space with the same boundary as $A$; also, the mutual information between two regions of space on the same flat boundary can be shown to be inversely proportional to the geodesic distance between the two regions as measured in the bulk AdS; moreover, the very connectivity between two regions of spacetime, thus the topology of the latter, has been conjectured to be related to the entanglement between the quantum degrees of freedom associated to the two regions \cite{vanramsdoonk}. This type of results suggest the possibility that entanglement measures, mutual information and the like can be used to {\it define} geometric and topological quantities and {\it reconstruct} spacetime out of the quantum properties of a many-body system that, a priori, is not about spacetime and geometry.   

\subsection{Quantum gravity many-body systems}
The above suggestion may sound radical and rather speculative. However, from the point of view of background independent quantum gravity approaches it is basically a necessity \cite{SN-entanglement}. The reason is the point we emphasized from the beginning: modern quantum gravity approaches are based on fundamental entities that are not gravitational or geometric per se, at least not in the standard GR sense of corresponding to continuum metric and matter fields, but are of more abstract, discrete quantum nature. Spacetime with its geometric properties has to emerge from their collective quantum dynamics. For this simple reason, we have anticipated, modern quantum gravity approaches are importing ideas and tools from condensed matter physics and many-body quantum systems. In fact, these formalisms, in their different ways, all propose a description of the microstructure of spacetime as a quantum many-body system. Let us see three examples.

As we mentioned, Loop Quantum Gravity \cite{LQG} (and its covariant version, spin foam models \cite{SF}) replace continuum manifolds and fields with purely combinatorial and algebraic data, with quantum states of space associated to (linear combinations of) spin networks: graphs labeled by irreducible representations of the rotation (or Lorentz) group. In turn, the graphs underlying such quantum states are dual to lattices and the algebraic data define for them a quantum geometry that reduces, in a classical limit (and under additional conditions), to a piecewise-flat discrete geometry (i.e. one in which each fundamental cell of the lattice is flat in the interior, with non-trivial curvature associated to their gluings and fully captured by the specification of lengths of the links of the lattice). Looking at such quantum states in more detail, one realises that they define rather peculiar lattice spin systems, generalised to arbitrary $SU(2)$ representations assigned to the lattice links and without any fixed lattice spacing (since the geometry of the lattice is to be determined by the dynamical algebraic data assigned to it. The quantum dynamics of the theory generically modified the combinatorial structure of the lattice itself and, possibly, its topology. Still, the similarity in mathematical structures with lattice many-body systems is striking, in particular they are very close to string-net condensate models, to spin chains and Ising systems, and of course to lattice gauge theories (which basically share the same type of state space). This implies on the one hand that techniques from quantum gravity have been used to study such statistical systems\cite{SN-ising}, but also that tools developed for them have been imported in quantum gravity, in particular for the study of coarse graining and continuum limits, i.e. to investigate the macroscopic phases of such quantum gravity models and the emergence of geometry \cite{bianca-renorm}. Tensor network renormalization methods, in particular, seem very promising \cite{SF-tensornetworks, SN-tensornetworks}. 

Another example is provided by random tensor models \cite{tensors}, a generalization to higher dimensions of matrix models for 2d quantum gravity and worldsheet string theory. The basic dynamical variables of these models are abstract rank-d tensors over N-dimensional index spaces to which one assigns a non-gaussian probability distribution depending on a number of coupling constants (they are themselves statistical systems). The connection to quantum gravity arises because their perturbative expansion generates a sum over lattices of topological dimension equal to $d$, with purely combinatorial amplitudes given as a path integral for gravity discretized on the same lattices interpreted as equilateral. The latter sum actually defines lattice quantum gravity formalisms like (euclidean) dynamical triangulations \cite{CDT}. Thus random tensor models attempt a purely combinatorial definition of the dynamical microstructure of spacetime, in terms of a statistical system of abstract tensors, in turn equivalent to a theory of random lattices. They have been developed to a great extent in rigorous mathematical terms as statistical systems, with many important results, like a well-defined large-N limit, double-scaling limits, some control over their critical behaviour, etc. They have been quite influential also in giving a solid mathematical basis for other quantum gravity formalisms, like group field theories, but also found some applications in the study of other statistical systems, like the Ising model on a random lattice or dimer models. More recently, as mentioned, they became central in yet another quantum gravity context, i.e. the AdS/CFT correspondence, by capturing the dynamics of the SYK model \cite{tensors-SYK}.

Finally, the already cited group field theories \cite{GFT} generalise random tensor models by replacing their finite index set with a Lie group, but maintaining the combinatorial structures defining them. By doing so, the models become bona fide quantum field theories, although the retain their combinatorial peculiarities, like the fact that the Feynman processes of the models are dual to d-dimensional lattices rather than graphs. The other crucial point is that, when the Lie group used is the rotation or the Lorentz group, their Hilbert space of states is a Fock space whose generic states are (linear superpositions of) spin networks, the same quantum states of Loop Quantum Gravity \cite{GFT2nd}, while their Feynman amplitudes are spin foam models (defining the covariant dynamics of spin networks), which can then be shown to be equivalent to discrete gravity path integrals (generalizing the ones of tensor models and dynamical triangulations to the case of dynamical edge lengths). Therefore, they naturally share the connection to quantum many-body and statistical systems that we have highlighted for these related formalisms. The new aspect, however, is that now quantum spacetime is tentatively described in a form that is {\it literally} a quantum many-body system, although with a more abstract physical interpretation (and of course not defined on any physical spacetime, with the base group manifold playing a different role) with a Fock space of states, a rather conventional field-theoretic definition of the quantum dynamics, etc. This allowed to use many of the conventional tools, suitably adapted, for tackling outstanding quantum gravity problems. Most notably, GFT renormalization \cite{GFTrenorm} is by now a thriving area of research, aiming establishing the renormalizability of GFT models for quantum gravity, and their continuum phase diagram, looking for a geometric, spatiotemporal phase. In parallel, the analogy with standard condensed matter systems have suggested a new approach to the extraction of an emergent gravitational dynamics, and of cosmology in particular as the simplest example of it. A very fertile direction of research \cite{GFTcosmo} has started from the hypothesis that the universe can be understood as a quantum condensate of the GFT \lq atoms of space\rq (the GFT quanta, forming spin networks and constituting its Hilbert space) and its effective condensate hydrodynamics (the GFT analogue of the Gross-Pitaevskii equation of real Bose condensates) can be recast in terms of cosmological dynamics of macroscopic observables like the volume of the universe, the density of emergent scalar matter, etc. While we are clearly just at the beginning in the development of this research direction, and in exploring the radical suggestions and potential new insights it provides, a number of important results have already been obtained: the emergent cosmological dynamics has the correct classical limit given by the Friedmann equation, but replaces the big bang singularity with a bouncing scenario, in which the universe passes from a previous contracting phase to the current expanding one \cite{GFT-bouncing}; the same dynamics has then been shown to have other interesting features, like the possibility of an accelerated inflation-like phase of expansion of purely quantum gravity origin (no need for additional degrees of freedom) \cite{GFT-accelerated}, and it has recently been extended to a framework for studying cosmological perturbations \cite{GFT-perturbations}, which appear to have a naturally close-to-scale invariant power spectrum and small amplitudes, as required to match CMB observations. 

\

To conclude, the above examples show that we are at the beginning of a long road, shaped jointly by quantum gravity ideas, in turn coming from gravitational physics as well as high energy physics, and by concepts and methods from condensed matter theory and quantum many-body systems. Moreover, they provide support to the idea that the universe we live in, including its very foundations, i.e. its spatiotemporal properties, can itself be understood as a peculiar quantum many-body system. An admittedly radical, but certainly also fascinating hypothesis, which is only beginning to prove its fruitfulness.

\end{document}